\documentclass[aip,apl,twocolumn,groupedaddress]{revtex4}

\usepackage{amssymb}
\usepackage{amsmath}
\usepackage{graphicx}
\usepackage{natbib}

\begin{document}
\textheight = 63\baselineskip
\title{Radio frequency charge sensing in InAs nanowire double quantum dots}

\author{M. Jung}
\author{M. D. Schroer}
\author{K. D. Petersson}
\author{J. R. Petta}
\affiliation{Department of Physics, Princeton University, Princeton, NJ 08544, USA}
\affiliation{Princeton Institute for the Science and Technology of Materials (PRISM), Princeton University, Princeton,
New Jersey 08544, USA}
\date{\today}

\begin{abstract}
We demonstrate charge sensing of an InAs nanowire double quantum dot (DQD) coupled to a radio frequency (rf) circuit. We measure the rf signal reflected by the resonator using homodyne detection. Clear single dot and DQD behavior are observed in the resonator response. rf-reflectometry allows measurements of the DQD charge stability diagram in the few-electron regime even when the dc current through the device is too small to be measured. For a signal-to-noise ratio of one, we estimate a minimum charge detection time of 350 $\mu$s at interdot charge transitions and 9 $\mu$s for charge transitions with the leads.
\end{abstract}

%\pacs{PACS go here}

\maketitle
Sensitive charge and spin state readout is necessary for the use of single spins in semiconductors as quantum bits. The quantum point contact (QPC) has been successfully used as a sensitive charge detector in single dot and DQD devices fabricated from GaAs/AlGaAs two-dimensional electron gas (2DEG) systems \cite{field93,petta05,elzerman05,johnson05,hanson07}. The conductance through an electrostatically coupled QPC is a very sensitive probe of the electron occupancy in a nearby quantum dot, allowing measurements of the full counting statistics in the current through a quantum dot as well as spin readout via spin-to-charge conversion \cite{gustavsson06, petta05}. While QPCs have proven to be powerful tools in 2DEG systems, their integration with lower dimensional systems, such as nanowires, requires challenging fabrication processes \cite{shorubalko08}.

A simple alternative approach to using a mesoscopic charge detector is to measure the dispersive shift of a resonant microwave circuit coupled to the DQD. In this way it is possible to probe the mesoscopic admittance of the DQD which loads the resonator \cite{cottet11}. In particular, in the low frequency limit, the DQD exhibits a state dependent `quantum capacitance' which  is proportional to the curvature of the energy bands, with a maximum capacitance occurring at interdot charge transitions \cite{duty05, ota10}. Using this technique, the quantum capacitance of a DQD has recently been observed and used to distinguish both charge and spin states \cite{petersson10, chorley12}.

In this letter, we demonstrate dispersive measurements of a few-electron InAs nanowire DQD device coupled to a microwave resonator held at milli-Kelvin temperatures. We compare standard dc transport measurements with the resonator response, which is probed using rf reflectometry \cite{schoelkopf98,cassidy07,reilly07,muller10}. With the device configured as a single dot, we observe Coulomb diamonds in dc transport and rf reflectometry measurements. In the DQD regime, we clearly observe the DQD charge stability diagram in the response of the reflected microwave signal. For a signal-to-noise ratio of one, we estimate a minimum state detection time of 9 $\mu$s for charge transitions with the leads and 350 $\mu$s for interdot charge transitions.

Samples are fabricated by first defining an array of Ti/Au depletion gates on a high-resistivity oxidized Si substrate. The gates are 20 nm thick, 30 nm wide, and spaced at a 60 nm pitch \cite{fasth07, nadj-pergel10, schroer11}. Two wide side-gates are also defined and allow the transparency of the nanowire leads to be tuned independently of the DQD confinement potential. A 20 nm layer of SiN$_{x}$ is then deposited as a gate dielectric. Nanowires are grown on InAs$<$111$>$B substrates using a gold-colloid seeded vapor-liquid-solid growth process in a metal-organic chemical vapor deposition reactor \cite{schroer10}. The InAs nanowires are then dispersed in ethanol and deposited on the substrate. The typical diameter of nanowires selected for device fabrication is $\sim$ 50--60 nm, as measured by atomic force microscopy. A 50/70 nm stack of Ti/Au is thermally evaporated to define ohmic contacts immediately after the contact region is passivated using an ammonium polysulfide ((NH$_{4}$)$_{2}$S$_{x}$) etch solution \cite{suyatin07}.

\begin{figure} [tr!]
\begin{center}
		\includegraphics[width=\columnwidth]{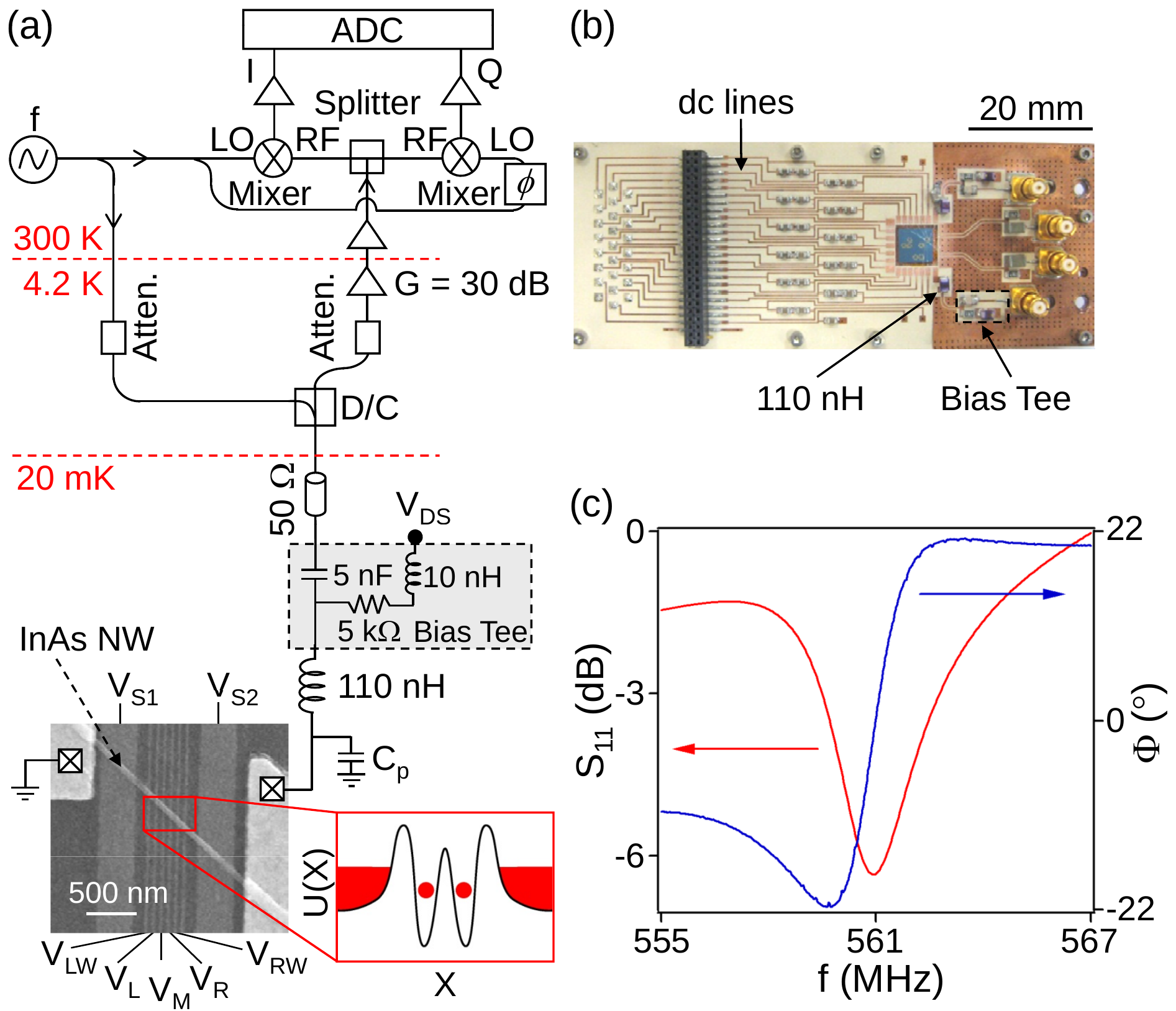}
\caption{\label{fig1} (Color online) (a) Circuit diagram of the measurement setup and scanning electron microscope (SEM) image of a representative InAs nanowire DQD device. (b) Photograph of the rf circuit board showing the sample, bias tee, and dc lines. (c) Amplitude and phase response measured using a network analyzer with the sample held at a temperature of 35 mK. }
\end{center}	
\end{figure}

High sensitivity rf charge detection is achieved by directly connecting the source electrode of the device to a resonant circuit consisting of a 110 nH chip inductor and its parasitic capacitance, $C_P$ $\sim$ 0.73 pF \cite{ind}. A circuit diagram is given in Fig.\ 1(a) and a picture of the circuit board is shown in Fig.\ 1(b). The circuit board has an on-board bias tee, enabling simultaneous rf and dc measurements in a dilution refrigerator with a base temperature $T$ $\sim$ 35 mK. Under typical operating conditions, the rf power incident on the resonant circuit is $\sim$ -85 dBm. The incoming and outgoing microwave signals are separated using a directional coupler with the reflected signal fed into a homemade cryogenic amplifier held at 4.2 K \cite{dc,weinreb09}. The signal is further amplified at room temperature and demodulated by mixing with the reference signal, allowing simultaneous measurements of the amplitude and phase response \cite{amp}. Figure 1(c) shows the amplitude and phase response of the rf circuit as a function of drive frequency $f$. The resonance frequency of the oscillator occurs at $f_{0}$ $\sim$ 561 MHz with a loaded quality factor $Q$ $\sim$ 140.

\begin{figure} [tr!]
\begin{center}
		\includegraphics[width=\columnwidth]{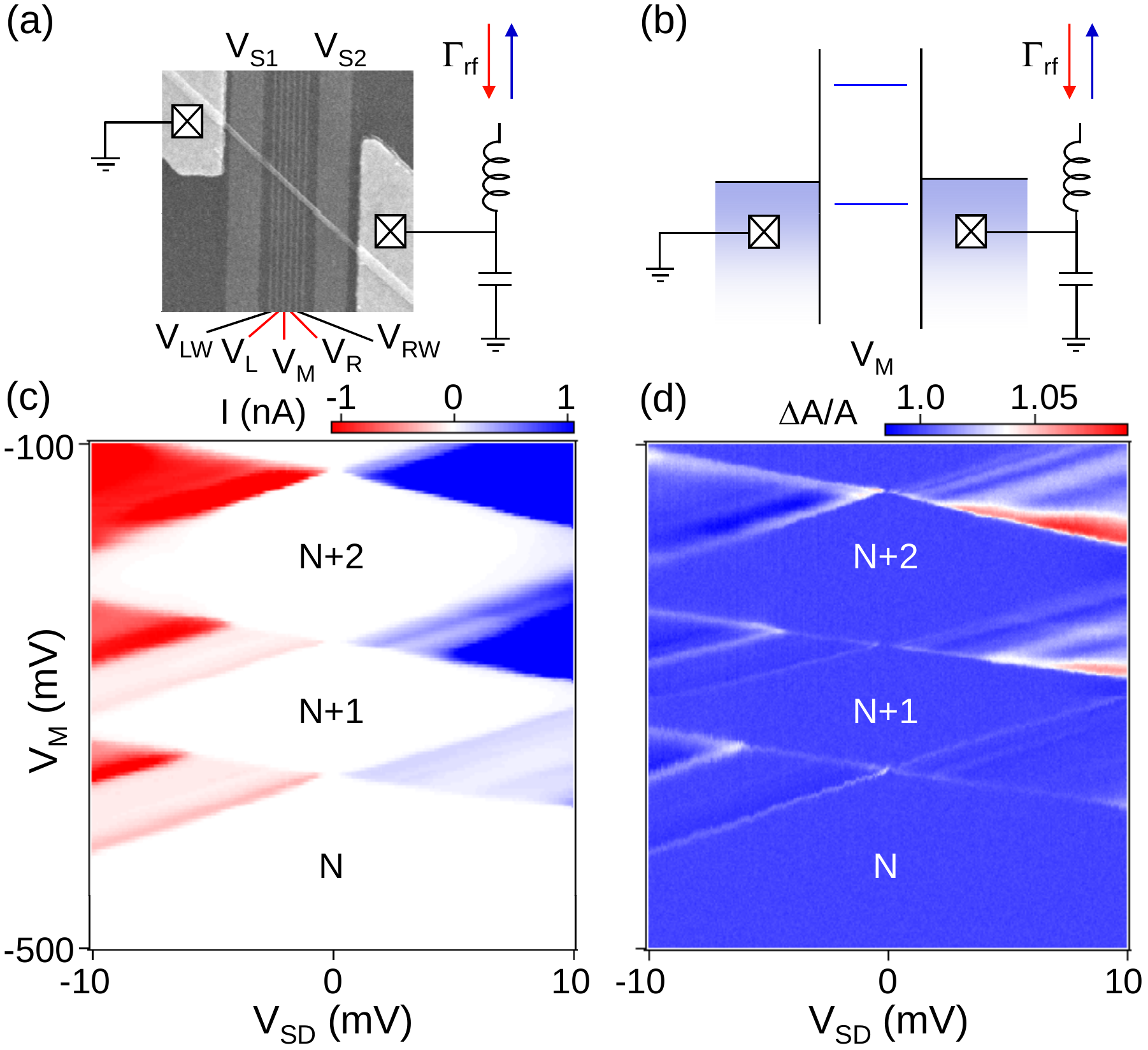}
\caption{\label{fig2} (Color online) (a) SEM image of a typical device. Three gates are energized with voltages $V_{L}$, $V_{M}$ and $V_{R}$ to define a single dot. Other gates are biased at positive voltages to prevent the formation of unintentional quantum dots. (b) Energy level diagram of the single dot. The gate voltage $V_{M}$ primarily controls the occupancy of the quantum dot, while $V_R$ and $V_L$ tune the height of the tunnel barriers. (c) Current through the dot, $I$, and (d) normalized amplitude response of the resonator, $\Delta A/A$, plotted as a function of $V_{M}$ and source-drain voltage $V_{SD}$. Coulomb diamonds are observed in both measurements.}
\end{center}	
\end{figure}

We first demonstrate measurements of a single quantum dot defined using three gates labeled $V_{L}$, $V_{M}$ and $V_{R}$ in Fig.\ 2(a) \cite{cheong02}. The other gates are held at positive bias to prevent the unintentional formation of quantum dots in regions of the wire leading to the source and drain contacts. A single dot energy level diagram is shown in Fig.\ 2(b). The left and right tunnel barriers are controlled with gate voltages $V_{L}$ and $V_{R}$, while $V_{M}$ controls the occupancy of the quantum dot. The current and resonator response of the single dot are measured as a function of source-drain voltage $V_{SD}$ and $V_{M}$ in Figs.\ 2(c)-(d), and clearly show Coulomb diamonds. Inside a Coulomb diamond, the number of electrons, $N$, trapped in the quantum dot is well defined due to the large electrostatic charging energy, $E_{C}$ = $e^2$/$C$, where $e$ is the electronic charge and $C$ is the total capacitance of the dot \cite{kouwenhoven01}. From the size of the $N$+1 Coulomb diamond we obtain a charging energy, $E_{C}$ $\sim$ 12 meV, which gives a dot capacitance $C$ $\sim$ 13 aF.

We next define a DQD using the same nanowire device with five gate voltages $V_{LW}$, $V_{L}$, $V_{M}$, $V_{R}$, and $V_{RW}$. The energy level diagram of the DQD is shown in the inset of Fig.\ 3. Interdot tunnel coupling is controlled using gate voltage $V_{M}$, while the tunneling rate to the drain (source) is adjusted with the gate voltages $V_{LW}$ ($V_{RW}$). Electron occupancy is tuned by the gate voltages $V_{L}$ and $V_{R}$. Figure 3 displays a large scale plot of the measured phase shift, $\Delta \Phi$, as a function of $V_{L}$ and $V_{R}$. A similar result is obtained in the amplitude response (see Fig.\ 4). Charge transitions appear as negative shifts in the phase of the reflected signal. The phase response reveals a DQD charge stability pattern, with charge stability islands labeled ($N_L$, $N_R$), where $N_L$($N_R$) is the number of electrons in the left(right) dot \cite{wiel02}. No additional charge transitions are observed in the lower left corner of the charge stability diagram, indicating a completely empty DQD, denoted (0, 0).

\begin{figure} [tr!]
\begin{center}
		\includegraphics[width=\columnwidth]{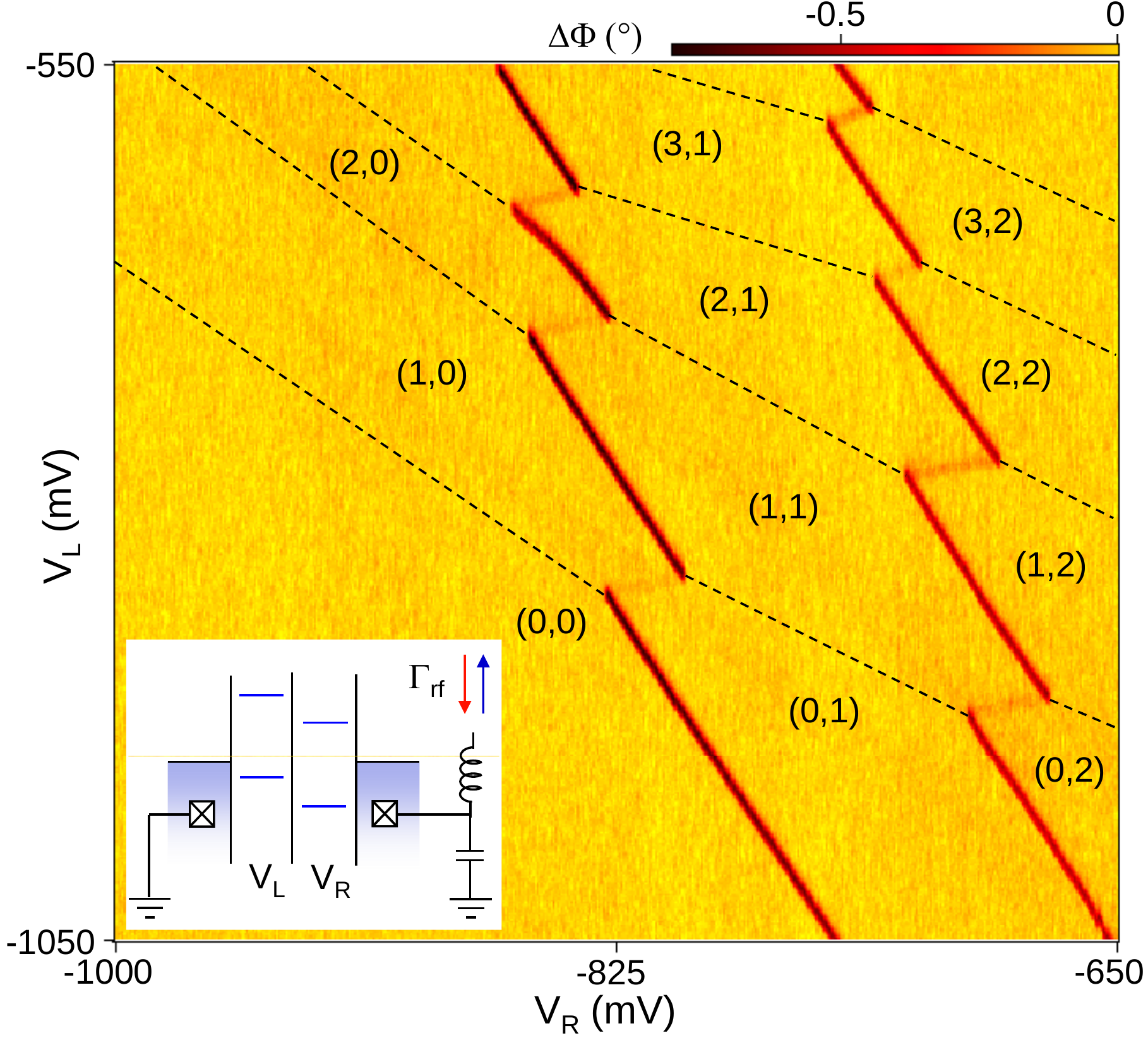}
\caption{\label{fig3} (Color online) Phase shift, $\Delta \Phi$, measured as a function of $V_R$ and $V_L$, with a source-drain voltage $V_{SD}$ = 0 V. The DQD charge stability diagram is clearly visible in the phase response of the resonator. Inset: Energy level diagram of the DQD. Gate voltages $V_L$ ($V_R$) primarily tune the occupancy in the left (right) dot.}
\end{center}	
\end{figure}

In Fig.\ 3 we readily observe charge transitions that change the electron number in the right dot, e.g. $(N_L, N_R)\leftrightarrow(N_L, N_R + 1)$. These charge transitions occur between the right dot and right lead to which the resonator circuit is connected. Given a tunnel rate $\gamma_R \gg 2\pi f_0$, the DQD has an effective capacitance along these transitions that is given by
\begin{equation}
C_{R} = \frac{\alpha_{R}^2 e^2 }{4k_B T},
\end{equation}
where $\alpha_{R} e$ is the lever arm that equates shifts in resonator voltage (at the source electrode) to changes in the right dot level relative to the Fermi energy of the right lead \cite{chorley12}. The effective capacitance adds to the total resonator capacitance and shifts the resonant frequency, thereby modifying the reflected signal. We note that in the case $\gamma_R \sim 2\pi f_0$, the effective capacitance of the DQD is reduced and out of equilibrium tunneling will result in energy dissipation with the DQD having an effective resistance that will damp the resonator. In contrast, when $\gamma_R \ll 2\pi f_0$ the resonator is largely insensitive to tunnel transitions. For this data set the phase response is largely insensitive to left dot charge transitions. The reduced response is due to both the lever arm $\alpha_{L}$ and tunnel rate $\gamma_L$ being much smaller for such transitions. In this particular sample, $\gamma_L$ could not be made sufficiently high to measure transport through the DQD while in the few-electron regime. It is possible that charge sensitivity could be increased by instead coupling one of the gate electrodes to the resonant circuit \cite{frey12}.

\begin{figure} [tr!]
\begin{center}
		\includegraphics[width=\columnwidth]{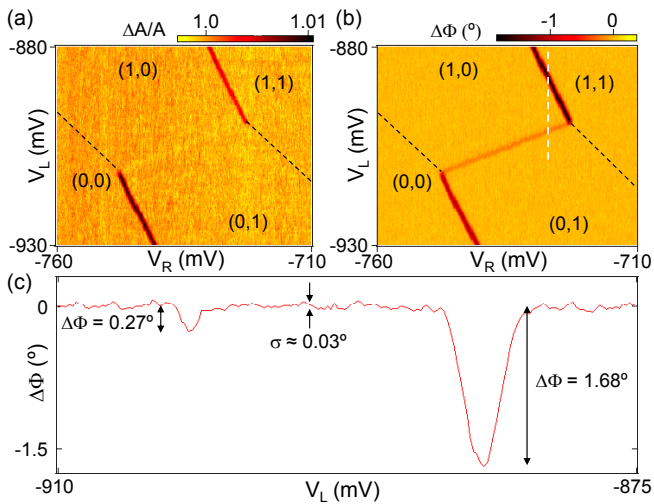}
\caption{\label{fig4} (Color online) (a) Normalized amplitude response, $\Delta A/A$, and (b) phase shift $\Delta \Phi$, measured as a function of $V_R$ and $V_L$. (c) $\Delta \Phi$ measured along the vertical dashed line in (b) reveals the (1,1) $\leftrightarrow$ (1,0) charge transition with the lead as well as the (1,0) $\leftrightarrow$ (0,1) interdot charge transition.}
\end{center}	
\end{figure}

In singlet-triplet spin qubits, spin-to-charge conversion is used for spin state readout. For example, a (1, 1) triplet state cannot tunnel to the (0, 2) singlet state due to the Pauli exclusion principle. However, a (1, 1) singlet state can tunnel to the (0, 2) singlet state. Spin-dependent interdot tunneling, combined with charge detection, is sufficient for spin state readout \cite{petta05}. It is therefore important to have sensitive charge detection along interdot charge transitions e.g., $(N_L+1, N_R)\leftrightarrow(N_L, N_R+1)$. We focus on the visibility of the phase and amplitude response at the one-electron interdot transition after a slight retuning of the device to increase the tunnel rates.

For the amplitude response data shown in Fig.\ 4(a), the interdot transition can barely be resolved. However, the phase data exhibit a robust response, attributable to the quantum capacitance of the double dot, which is given by \cite{petersson10, duty05},
\begin{equation}
C_Q = \frac{\alpha_{\epsilon}^2e^2}{4t},
\end{equation}
where $t$ is the interdot tunnel coupling and $\alpha_{\epsilon} e$ is the lever arm that relates the resonator voltage to changes in the detuning of the left and right dot levels. This capacitance shifts the resonant frequency which, in turn, modifies the phase response. We expect that for $C_Q \ll C_P$, the amplitude response at the resonance is first order insensitive to such frequency shifts, consistent with Fig.\ 4(a). Figure 4(c) shows the phase response as a function of gate voltage $V_{L}$ taken along a cut indicated by the dashed vertical line in Fig.\ 4(b). Phase shifts are measured to be 0.27 degrees at the interdot transition and 1.68 degrees at the (1, 1) $\leftrightarrow$ (1, 0) transition. The measured phase shifts at resonance can be related to the quantum capacitance $\Delta\Phi$ $\sim$ 2$Q$ $\times$ $C_Q/C_P$ \cite{cassidy07, pozar05}. We estimate a capacitance change of 76 aF at the charge transition with the source electrode. Likewise, along the (1, 0) $\leftrightarrow$ (0, 1) interdot charge transition, the capacitance change is estimated to be $\sim$ 12 aF. With a lever arm of $\sim 0.15$ eV/V, this gives an interdot tunnel coupling $t \sim 75$ $\mu$eV.

We now consider the sensitivity of this detection technique by comparing the background phase noise with the signal obtained at interdot and single dot charge transitions. From the data in Fig.\ 4(c), we measure a root-mean-square noise amplitude of $\sigma \approx 0.03$ degrees with an effective bandwidth of $B = 35$ Hz (7 kHz low-pass filter and 200 line averages). Based on the phase shift at a charge transition, we can estimate the detection time for a signal-to-noise ratio of one using $\tau_{min} = (\sigma/\Delta\Phi)^2/B$. For interdot and single dot charge transitions we extract detection times of $\tau_{min} \sim$ 350 ${\mu}$s and $\tau_{min} \sim$ 9 ${\mu}$s respectively. The detection time for the interdot response sets a minimum time for distinguishing ground and excited charge states, due to the difference in sign of the quantum capacitance at zero detuning \cite{petersson10}. Furthermore, two-electron singlet and triplet spin states can also be measured on a comparable timescale as the quantum capacitance of the triplet state is zero near the interdot charge transition due to Pauli blockade \cite{petta05,petersson10}.

Acknowledgements: Research at Princeton was supported by the Sloan and Packard Foundations, Army Research Office grant W911NF-08-1-0189, DARPA QuEST award HR0011-09-1-0007 and the National Science Foundation through the Princeton Center for Complex Materials, DMR-0819860 and CAREER award, DMR-0846341. Research was carried out in part at the Center for Functional Nanomaterials, Brookhaven National Laboratory, which is supported by DOE BES Contract No. DE-AC02-98CH10886. Partially sponsored by the United States Department of Defense. The views and conclusions contained in this document are those of the authors and should not be interpreted as representing the official policies, either expressly or implied, of the U.S. Government.

%\bibliographystyle{aipnum4-1}
%\bibliography{references}

%
\end{document}